\documentstyle[preprint,prl,aps]{revtex}
\begin{document}
\title{Universality in the One-Dimensional Self-Organized Critical 
Forest-Fire Model}
\author{Barbara Drossel, Siegfried Clar, and Franz Schwabl}
\address{Institut f\"ur Theoretische Physik, \\
         Physik-Department der Technischen Universit\"at M\"unchen, \\
         James-Franck-Str., D-85747 Garching, Germany}
\date{\today}
\maketitle
\begin{abstract}
We modify the rules of the self-organized critical forest-fire model in one
dimension by
allowing the fire to jump over holes of $\le k$ sites.
An analytic calculation shows that not only the size distribution of forest
clusters but also the size distribution of fires is characterized by the same
critical exponent as in the nearest-neighbor model, i.e. the critical behavior
of the model is universal. Computer simulations confirm the analytic results.
\end{abstract}
\pacs{PACS numbers: 05.40.+j, 05.70.Jk, 05.70.Ln}


\narrowtext
\section{Introduction}
\label{u1}

Some years ago, Bak, Tang, and Wiesenfeld introduced the {\em sandpile model}
which
evolves into a critical state irrespective of initial conditions and
without fine tuning of para\-meters \cite{bak1}. Such systems are called {\em
self--organized critical} (SOC)  and exhibit power--law correlations in space
and time. The concept of SOC has attracted much
interest since it might explain the origin of {\em fractal structures} and
{\em $1/f$--noise}. Other SOC models e.g. for earthquakes
\cite{chr1,che} or the evolution
of populations \cite{bak2,fly1} have been introduced since then, improving our
understanding of the mechanisms leading to SOC.
Recently, a forest-fire model has been introduced which can be viewed as a
model 
for excitable media \cite{tys,mer}. It becomes self-organized
critical when time scales are separated \cite{dro1}.
In one dimension, the critical exponents could be determined analytically, thus
proving the possibility of SOC in nonconservative systems \cite{dro2}.

Analogous to critical phenomena in equilibrium phase transitions, it is
expected that the values of the critical exponents depend only on few
macroscopic properties of the system as dimension, conservation laws and
symmetries, i.e. the critical behavior of the model is {\em universal}.
Computer simulations of the forest-fire model for different lattice symmetries
and for a modification with immune trees 
show indeed universal behavior \cite{cla1,dro5}, but so far this observation
has no analytic foundation.

In this paper, we show by analytic means that the critical exponents in the
one-dimensional forest-fire model are universal when the fire is allowed to
jump over holes up to a given size. In Sec. \ref{u2}, we introduce the rules
of the model. 
In Sec. \ref{u3}, we give a short review of the analytic solution of Ref.
\cite{dro2}. 
In Sec. \ref{u4}, we calculate the critical exponent for the fire size
distribution when the fire is allowed to jump over holes of up to $k$ sites.
The values of the exponents are confirmed by computer simulations. Finally, we
summarize  our results.

\section{The model}
\label{u2}

The forest-fire model is a stochastic cellular automaton which is defined on a
hypercubic lattice with  $L^d$ sites. In this paper, we consider only the
one-dimensional case $d=1$. Each site
is occupied by
a tree, a burning tree, or it is empty. During one time step, the system is
parallely updated according to the following rules
\begin{itemize}
\item burning tree $\longrightarrow$ empty site
\item tree $\longrightarrow$ burning tree, 
if at least one neighbor in a distance $\le k+1$ is burning,  $k=0,1,2,
\dots$
\item tree $\longrightarrow$ burning tree with probability $f$,  
if no neighbor is burning
\item empty site $\longrightarrow$ tree with probability $p$.
\end{itemize}

Starting with arbitrary initial conditions, the system
approaches after a transition period a steady state the
properties of which depend only on the parameter values. We always assume that
the lattice is so large that no finite size effects occur.
The steady state is self-organized critical if the parameters satisfy a double
separation of time scales
\begin{equation}
f\ll p \ll f/p \, .\label{eq1}
\end{equation}
The first inequality means that many trees grow between two lightning
strokes and therefore large forest clusters and fires occur. The second
inequality means that even 
large forest clusters burn down before new trees grow at their edge.
Under these conditions, the size distributions of forests and fires obey power
laws as we shall see below. 

\section{Analytic solution for $ \lowercase{k} = 0 $}
\label{u3}

In the case $k=0$, where the fire is stopped by any empty site, i.e. just jumps
to nearest neighbors, many properties
of the model have been derived analytically in \cite{dro2}.
Before proceeding to general $k$, we give an illustrative derivation of these
results.  

The mean number of trees destroyed by a lightning stroke is
\begin{equation}
\bar s=(f/p)^{-1}(1-\rho)/\rho,
\label{eq2}
\end{equation}
where $\rho$ is the mean forest density in the steady state \cite{dro1}.

Let $n(s)$ be the mean number of forest clusters of $s$ trees, divided by the
number of sites $L$. $n(s)$ will be shown to obey a power law  
\begin{equation}
n(s)\propto s^{-\tau} \label{eq3}
\end{equation}
for clusters smaller than a cutoff
\begin{equation}
s_{\text{max}} \propto (f/p)^{-\lambda}, \label{eq4}
\end{equation}
besides of possible logarithmic corrections.
The probability that lightning strikes a forest cluster of size $s$ is
proportional to $sn(s)$. Since the fire is stopped by any empty site, the size
distribution of fires is also proportional to $sn(s)$.

In order to derive the size distribution of fires, consider a string of $n \ll
p/f$ sites. This string is too short for two trees to grow during
the same time step. Lightning does not strike this string before all of its
trees are 
grown. Since we are always interested in the limit $f/p \to 0$, the following
considerations remain valid even for strings of a very large  size. 
Starting with a completely empty state, the string passes
through a cycle which is illustrated in Fig.~\ref{fig1}. During one time step,
a tree grows with
probabilty $p$ on any site. After some time, the string is
completely occupied by trees. Then the forest in the neighborhood of our string
will also be quite dense. The forest on our string is part of a forest cluster
which is much larger than $n$. Eventually that cluster becomes so large that it
is struck by lightning with a nonvanishing probability. Then the forest cluster
burns down, and the string again becomes completely empty. (For a rigorous
justification of the neglection of lightning strokes on the n-string leading to
random growth of trees, we refer the reader to Ref.~\cite{dro2}.)

This consideration allows us to write down rate equations for the states of the
string. In the steady state, each configuration of trees is generated as often
as it is destroyed. Let $P_n(m)$ be the probability that our string is occupied
by $m$ trees. Each
configuration which contains the same number of trees has the same probability.
A configuration of $m$ trees is destroyed when a tree grows at one of the empty
sites, and is generated when a tree grows in a state consisting of $m-1$
trees. 
The completely empty state is generated when a dense forest burns down. Since
all trees on our string burn down simultaneously, this happens each time when a
given site of our string is on fire. This again happens as often as a new tree
grows at this given site, i.e. with probability $p\, (1-\rho)$ per time step.
We therefore have the following equations (which have been derived more
formally in 
\cite{dro2}) 
\begin{eqnarray*}
p n P_n(0) & = & p\, (1-\rho), \\
p(n-m)P_n(m) & = & p\, (n-m+1)P_n(m-1)\; 
\text{ for}\; m \neq 0,n\, .
\end{eqnarray*}
We conclude
\begin{eqnarray}
P_n(m) & = & (1-\rho)/(n-m)\; \text{ for}\;m<n, \nonumber \\
P_n(n) & = & 1-(1-\rho)\sum_{m=0}^{n-1}1/(n-m) \label{eq5} \\
& = & 1-(1-\rho) \sum_{m=1}^n 1/m. \nonumber
\end{eqnarray}

A forest cluster of size $s$ is a configuration of $s$ neighboring trees with
an empty site at each end. The size distribution of forest clusters
consequently is 
\begin{equation}
n(s) ={P_{s+2}(s)\over 
{s+2 \choose s}} =  {1-\rho\over (s+1)(s+2)} \simeq (1-\rho)s^{-2}.
\label{eq6}
\end{equation}
This is a power law with the critical exponent $\tau=2$.
The size distribution of fires is 
$\propto sn(s)\propto s^{-1}$.

There is a characteristic length $s_{\text{max}}$ 
where the power law $n(s)\propto s^{-2}$ breaks down. 
We calculate 
$s_{\text{max}}$ from the 
condition that a string of size $n
\le s_{\text{max}}$ is not 
struck by lightning until all trees are
grown. When a string of size $n$ is completely empty at time
$t=0$, it will be occupied by $n$ trees after
\begin{displaymath}
T(n)=(1/p)\sum_{m=1}^n 1/m \simeq \ln(n)/p
\end{displaymath}
 timesteps on an
average. The mean number of trees after $t$ timesteps is 
\begin{displaymath}
m(t)=n[1-
\exp(-pt)].
\end{displaymath}
 The probability that lightning
strikes a string of size $n$ before all trees are grown is
\begin{displaymath}
f\sum_{t=1}^{T(n)} m(t)\simeq (f/p)n(\ln(n)-1) \simeq
(f/p)n\ln(n).
\end{displaymath}
 We conclude 
\begin{equation}
s_{\text{max}}\ln(s_{\text{max}}) 
\propto p/f \; \text{ for large $p/f$}\, ,
\label{eq7}
\end{equation}
leading to $\lambda=1$.

Next we determine the relation between the mean forest density
$\rho$ and the parameter $f/p$. 
The mean forest density is given by
\begin{eqnarray*}
\rho & \simeq & \sum_{s=1}^{s_{\text{max}}} sn(s) \\
     & = & (1-\rho)\sum_{s=1}^{s_{\text{max}}}{s\over 
           (s+1)(s+2)}\\
     & \simeq & (1-\rho) \ln(s_{\text{max}}).
\end{eqnarray*}
Thus
\begin{equation}
{\rho \over 1-\rho}  \simeq  \ln(s_{\text{max}}) \simeq \ln(p/f)\, \; \text{
for large $p/f$}. 
\label{eq8}
\end{equation}
The forest density approaches the value 1 at the 
critical point. This is not
surprising since no infinitely large 
cluster exists in a one-dimensional
system as long as the forest is not completely dense.
Combining Eqs.~(\ref{eq6}) and (\ref{eq8}), we obtain the 
final result for the cluster-size distribution near the 
critical point
\begin{equation}
n(s) \simeq \frac{1}{(s+1)(s+2)\ln s_{\text{max}}} \;
\text{for} \; s < s_{\text{max}}
\end{equation}
with $s_{\text{max}}$ given by Eq.~(\ref{eq7}).

The size distribution $sn(s)$ of the fires has also been determined by computer
simulations. The result is shown in Fig.~\ref{fig2}. It agrees perfectly with
Eq.~(\ref{eq6}) in the region $s < s_{\text{max}}$. 

\section{Universality of the critical exponents}
\label{u4}

We now allow the fire to spread to trees up to a distance $k+1$ from a burning
tree, 
as given by the second rule above. The fire jumps over holes of up to $k$ empty
sites, but 
is stopped by holes of more than $k$ sites. Consequently a fire no longer
destroys just a single forest cluster, but it may also destroy several clusters
which are separated by holes of $\le k$ sites. The size distribution of fires
therefore is no longer given by $sn(s)$. In this section, we will show that the
critical exponent which describes the size distribution of fires is
 still $1$ indicating its universality.

We consider a string of $n \gg k$ sites which is empty in the beginning. After
$t$ timesteps, it contains a hole of size $k+1$ with the probability 
\begin{displaymath}
n\,(1-p)^{(k+1)t}(1-(1-p)^t)^2 \simeq n\, e^{-(k+1)tp}\, 
\end{displaymath}
for small $p$ but large $pt$. The time after which there are no holes larger
than $k$ 
therefore is proportional to $\ln(n)/p\,(k+1)$. This time becomes very long for
large values of $n$, and consequently the forest is very dense at the moment
where  all holes
larger than $k$ have disappeared. The critical forest density therefore is
still $\rho_c=1$. As long as $\rho<1$, there is a nonvanishing probability that
a hole larger than $k$ occurs, and the fire cannot spread indefinitely.

As in the previous section, we choose $n$ so small that the string is not
struck by lightning before all empty sites have disappeared, i.e.
$n<s_{\text{max}}$. In the limit $f/p\to 0$, $s_{\text{max}}$ diverges, and the
string can be very large. 
The dynamics on our string are exactly the same as
before (Fig.~\ref{fig1}), and Eqs.~(\ref{eq5}) to (\ref{eq7}) are still valid.
The size 
distribution of forest clusters smaller than $s_{\text{max}}$ remains
unchanged, and the critical exponents $\tau=2$ and $\lambda=1$ characterizing
this distribution 
are universal. Eq.~(\ref{eq8}) for the forest density also remains the same in
the limit $f/p \to 0$ (besides of a constant which has to be added to the
right-hand 
side of Eq.~(\ref{eq8}) but which has already been neglected before, since it
is much smaller than $\ln(p/f)$).

Now we calculate the size distribution $F(m)$ of fires that destroy $m$ trees.
For $m<s_{\text{max}}$, it can be derived using Eqs.~(\ref{eq5}). It is
proportional to $m$ times the number of configurations which contain $m$ trees
with at least $k+1$ empty sites at each end and no holes larger than $k$ sites
between the trees, i.e.
\begin{equation}
F(m) \propto m \sum_{N=0}^{k(m-1)} {P_{m+N+2k+2}(m)\over {m+N+2k+2\choose m}} 
\; \sum_{\{l_0,...,l_k\},\, \sum_i l_i=m-1, \, \sum_i il_i = N}\;
{(m-1)! \over l_0! \cdots l_k!} \, . 
\label{eq10}
\end{equation}
Since we are only interested in the asymtotic power law for large $m$, this sum
can be simplified. The main contribution comes from values $N \ll m$ since
for larger $N$ 
there are only few configurations which contain no holes larger than $k$ (see
the consideration at the beginning of this section). 
The probability for a hole of size $k+1$ on a string of $m+N$ sites with $N\ll
m$ empty sites is 
\footnote{This result is valid only for $N\ll m$ and is obtained as follows:
There are $m-1$ possible 
positions for the hole of size $k+1$. The remaining $N-k-1$ empty sites can be
distributed at random over the remaining $m-2$ gaps between trees. To obtain a
probability, we have to divide by all possible configurations of $N$ empty
sites on $m-1$ gaps.}
\begin{displaymath}
\simeq  (m-1) {(m-2)^{N-k-1} / (N-k-1)! \over (m-1)^N /N!} \simeq {N^{k+1}
\over m^k}\, , 
\end{displaymath}
from which we conclude that the first sum in Eq.~(\ref{eq10}) has a cutoff for
$N 
\propto m^{k/(k+1)}$. The second sum in Eq.~(\ref{eq10}) counts the number of
different configurations of $N$ empty sites on $m-1$ gaps with the restriction
that each gap contains no more than $k$ empty sites. For $N$ smaller than the
cutoff, the probability for holes larger than $k$ is very small anyway, and we
are therefore allowed to sum over all configurations of $N$ empty sites on
$m-1$ gaps which gives ${m+N-2 \choose m-2}$.
\footnote{This is equivalent to the number of configurations of $m$ trees on
$N+m$ sites with a tree at both ends.}
We then obtain
\begin{equation}
F(m) \propto \sum_{N=0}^{m^{k/(k+1)}} {m\over N+2k+2} {{m+N-2 \choose m-2}\over
{ m+N+2k+2 \choose m}} \simeq \sum_{N=0}^{m^{k/(k+1)}} \left({N\over m
}\right)^{2k+1} \propto {1\over m} \, . 
\label{eq11}
\end{equation}
This is a power law with an exponent $1$ which is independent of $k$. 
We thus have shown that not only the exponent for the size distribution of
forests but also the exponent for the size distribution of fires is universal.
The form of the  cutoff functions describing the behavior of these
distributions on lengths 
larger than $s_{\text{max}}$, however, is different for different values of
$k$. 

Our computer simulations confirm the analytic result. In Fig.~\ref{fig3} and
Fig.~\ref{fig4}, the size distribution of fires is shown for fire propagation
over holes of size 1 and 2. The slope in the scaling region is $-1$ each time. 

\section{Conclusion}
\label{u5}

In this paper we have shown by analytic means that the critical exponents of
the SOC 
forest-fire model in one dimension show universal behavior when the range of
the interaction is changed. This is additionally confirmed by computer
simulations. 

In two dimensions, too, computer simulations show  that the model is universal
under a change of the lattice symmetry. It still remains a challenge to prove
universality in dimensions higher than two analytically, e.g. by the
renormalization group formalism. 


\begin{figure}
\caption{Dynamics on a string of $n=4$ sites. Trees are black, empty sites are
white.}
\label{fig1}
\end{figure}

\begin{figure}
\caption{Size distribution of the fires for $f/p=1/25000$ and $L=2^{20}$.
The smooth line is the theoretical result, which is valid for cluster 
sizes $\le s_{\text{max}}$.}
\label{fig2}
\end{figure}

\begin{figure}
\caption{Size distribution of the fires for $f/p=1/8000$ and $ L=2^{20}$.
The fire is allowed to jump over holes of one empty site.}
\label{fig3}
\end{figure}

\begin{figure}
\caption{Size distribution of the fires for $f/p=1/8000$ and $L=2^{20}$.
The fire is allowed to jump over holes of two empty sites.}
\label{fig4}
\end{figure}


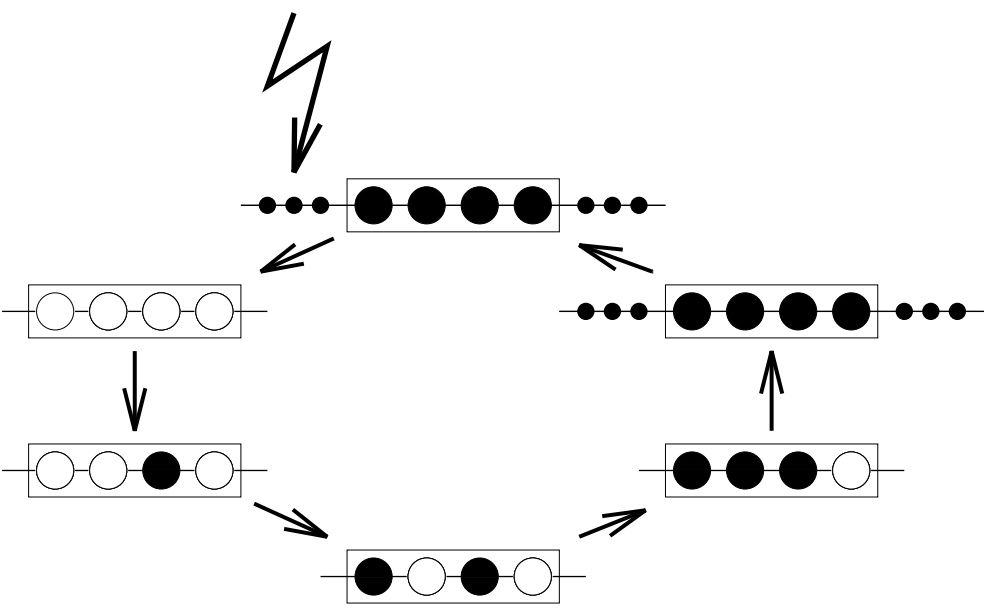
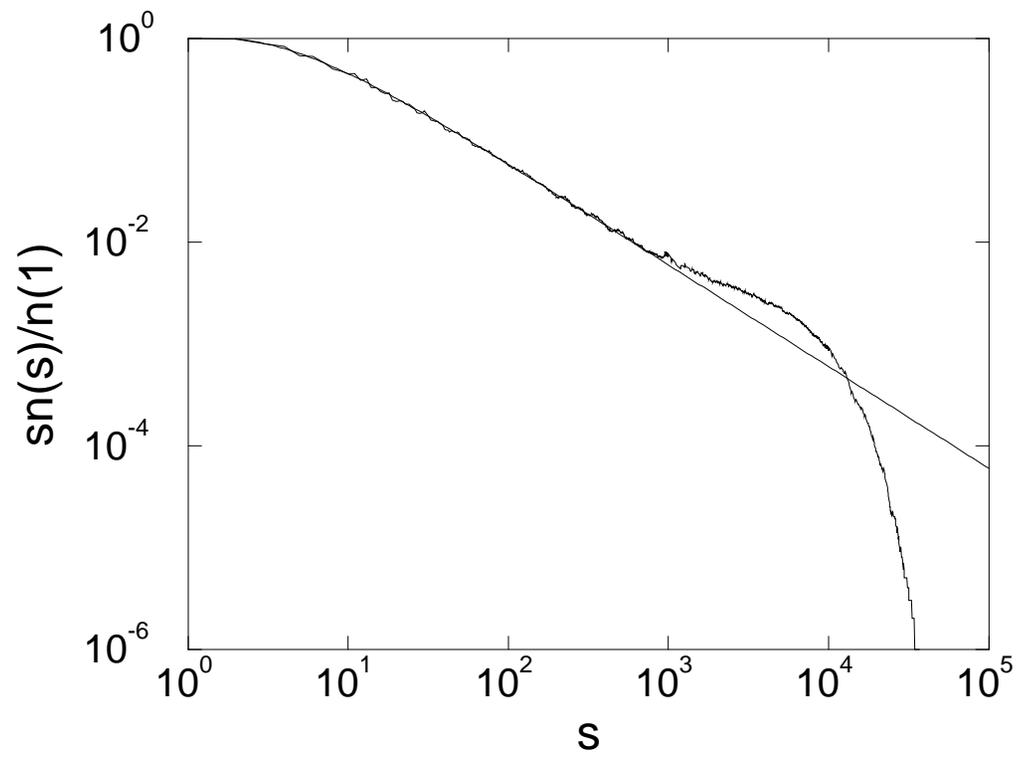
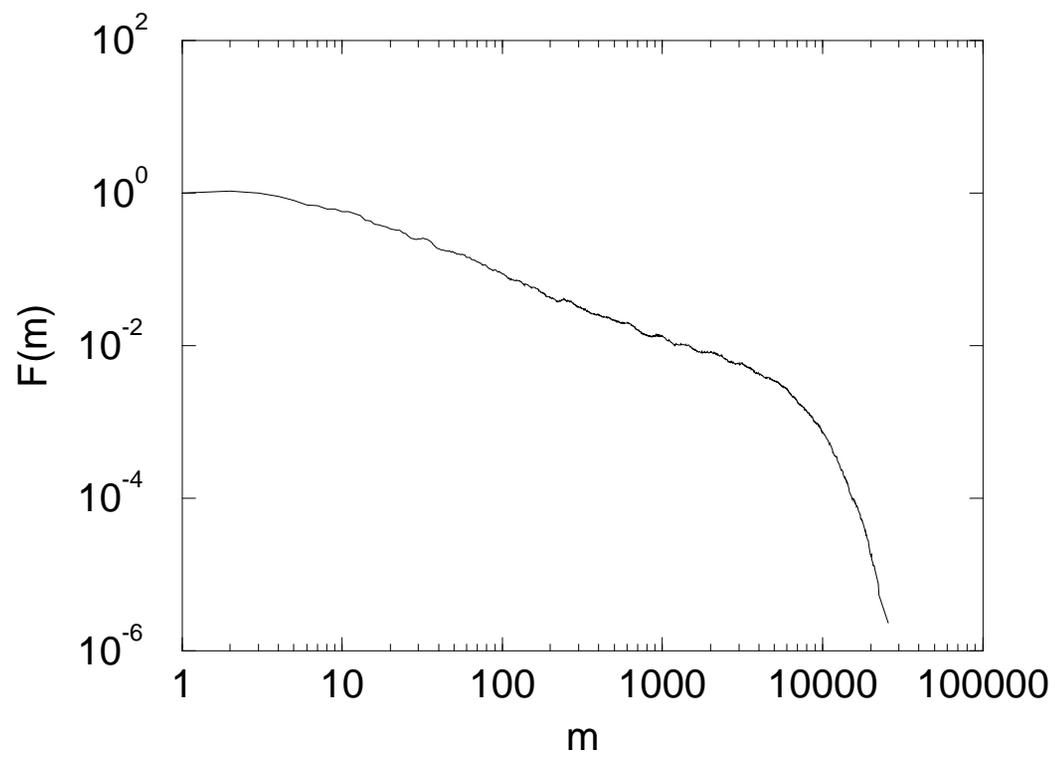
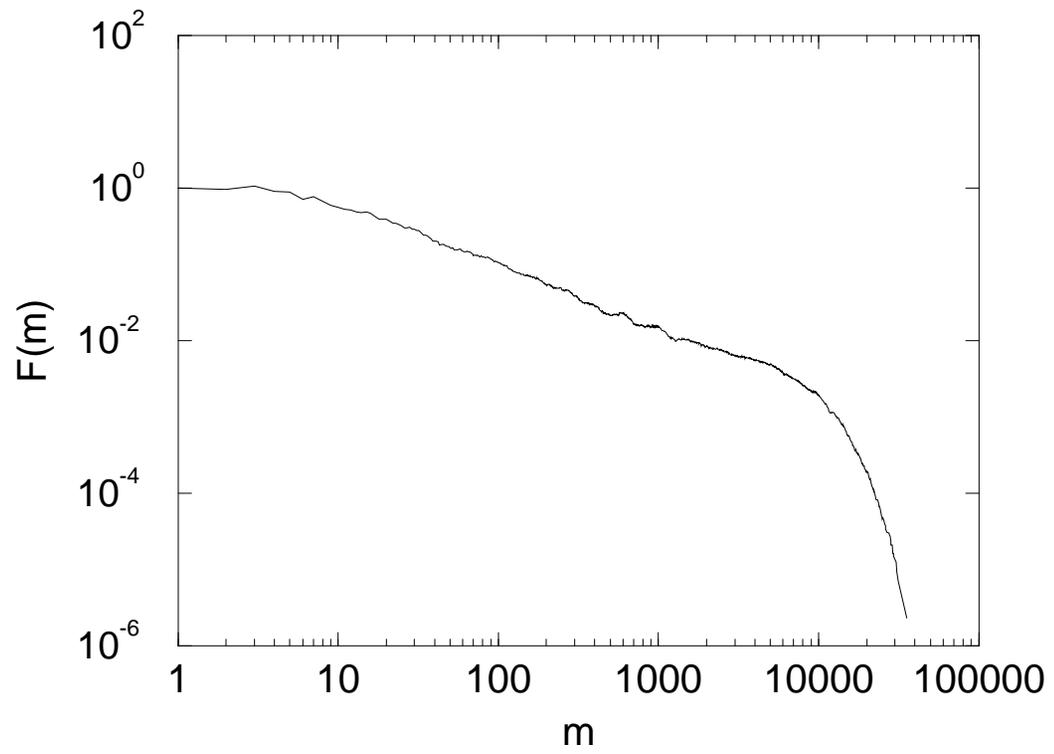
\end{document}